\documentclass[fleqn,usenatbib]{mnras}

\usepackage[usenames,dvipsnames]{xcolor}
\usepackage{amsmath}
\usepackage{array}
\usepackage{xfrac}
\usepackage{multirow}

\newcommand{\rev}[1]{{\textcolor{black}{#1}}}
\usepackage[T1]{fontenc}
\usepackage{ae,aecompl}
\hypersetup{draft}

\title[How do SFHs affect abundances?]{SDSS-IV MaNGA: how do star-formation histories affect gas-phase abundances?} 

\author[Boardman et al.]{
N.~Boardman$^{1}$\thanks{E-mail: nfb@st-andrews.ac.uk},
V.~Wild$^{1}$,
K.~Rowlands$^{2,3}$, 
N.~Vale Asari$^{4}$,
Y.~Luo$^{3}$\\
$^{1}$School of Physics and Astronomy, University of St Andrews, North Haugh, St Andrews KY16 9SS, UK\\
$^{2}$AURA for ESA, Space Telescope Science Institute,
3700 San Martin Drive, Baltimore, MD 21218, USA\\
$^{3}$William H. Miller III Department of Physics and Astronomy, Johns Hopkins University, Baltimore, MD 21218, USA\\
$^{4}$Departamento de F\'{\i}sica--CFM, Universidade Federal de Santa Catarina, C.P.\ 476, 88040-900, Florian\'opolis, SC, Brazil \\
}

\date{Accepted X. Received X; in original form X}
\pubyear{2022}
\begin{document} 
\label{firstpage}
\pagerange{\pageref{firstpage}--\pageref{lastpage}}
\maketitle

\begin{abstract}

Gas-phase abundances in galaxies are the products of those galaxies’ evolutionary histories. The star-formation history (SFH) of a region might therefore be expected to influence that region’s present day gaseous abundances. Here, we employ data from the MaNGA survey to explore how local gas metallicities relate to star-formation histories of galaxy regions. We combine MaNGA emission line measurements with SFH classifications from absorption line spectra, to compare gas-phase abundances in star-forming regions with those in regions classified as starburst, post-starburst and green valley. We find that starburst regions contain gas that is more pristine than in normal star-forming regions, in terms of O/H and N/O; we further find that post-starburst regions (which have experienced stochastic SFHs) behave very similarly to ordinary star-forming regions (which have experienced far smoother SFHs) in O/H--N/O space. We argue from this that gas is diluted significantly by pristine infall but is then re-enriched rapidly after a starburst event, making gas-phase abundances insensitive to the precise form of the SFH at late times. We also find that green-valley regions possess slightly elevated N/O abundances at a given O/H\rev{; this is potentially due to a reduced star-formation efficiency in such regions, but it could also point to late-time rejuvenation of green valley regions in our sample}.

\end{abstract}


\section{Introduction}\label{intro}

Gas-phase abundances in galaxies are the products of galaxies’ evolutionary histories. Present-day abundances are likely shaped by competing physical processes such as pristine inflow, star-formation and metal-rich outflow \citep[e.g.][]{schmidt1963,lilly2013,zhu2017,bb2018}, with successive generations of stars serving to enrich gas over time. In such a scenario, gas-phase abundances should be entwined with star-formation histories (SFHs) on a range of scales. Integral-field-unit (IFU) spectroscopic surveys — such as CALIFA \citep{sanchez2012}, SAMI \citep{croom2012}  and MaNGA \citep{bundy2015} — enable detailed mapping of galaxies’ stellar and gaseous contents over a two-dimensional field, making them ideal to study how SFH connects to gas-phase abundances across galaxies. Correspondingly, a large number of IFU studies have considered two-dimensional maps of galaxies' chemical abundances \citep[e.g.][]{sanchez2012b,stott2014,belfiore2017,poetrodjojo2018}, with such studies serving as natural extensions from work involving single fibre spectra \citep[e.g.][]{tremonti2004,ellison2008} or long-slit spectra \citep[e.g.][]{lequeux1979,ferguson1998,vanzee1998,moustakas2010}. 

Observational gas-phase abundance studies typically focus on one or both of the following: the gas-phase metallicity $\mathrm{12 + \log(O/H)}$ and the nitrogen-to-oxygen ratio $\mathrm{\log(N/O)}$. The two parameters provide complementary information, due to the different timescales and physics that they probe: $\alpha$-elements such as oxygen are \textit{primary} elements produced via nucleosynthesis in massive stars and then released into the ISM via core-collapse supernovae, which occur in massive stars on roughly $\sim$ 10Myr timescales after a star-formation episode \citep[e.g.][]{timmes1995}. Nitrogen by contrast is both a \textit{primary} and a \textit{secondary} element \citep[e.g.][]{matteucci1986}; it is released into the ISM on $\sim$ Gyr timescales after a star-formation event \citep[e.g.][Figure 1, attributed to Vincenzo et al. in prep]{maiolino2019}, in quantities that depend on the initial abundance of both oxygen and carbon. Observationally, O/H and N/O are quite tightly correlated at high metallicities, with N/O displaying an approximately constant value at low metallicities \citep{edmunds1978,izotov1999,molla2006,berg2012,izotov2012,andrews2013,james2015}.

On galaxy scales, both O/H and N/O have been found to correlate significantly with stellar mass $\mathrm{M_\star}$ \citep[e.g.][]{tremonti2004,pm2013,masters2016}, with the former relation typically referred to as the mass--metallicity relation (MZR). The MZR has been found to display a residual dependence on star-formation rate (SFR), with higher SFRs associated with lower metallicities at a given mass, producing a three-way relation known as the fundamental metallicity relation \citep[FMR; e.g.][]{ellison2008,ll2010,mannucci2010}. On the other hand, the FMR's reduction in scatter from the MZR appears to be relatively modest and is seen mostly in lower-mass ($\mathrm{M_\star < 10^{10} M_\odot}$) galaxies \citep[e.g.][]{salim2014}, which are potential factors in the reported non-detections of a galaxy-scale FMR in smaller IFU data-sets \citep[e.g.][]{sanchez2013,sanchez2017a,bb2017, ah2022}. \citet{cresci2019} have investigated this point for CALIFA and MaNGA data, and they argue that a galaxy-scale FMR remains apparent when the relation's mass dependence is considered. 

\rev{For N/O, measurements typically support little residual SFR dependence, resulting in a particularly tight trend between N/O and $\mathrm{M_\star}$ \citep[e.g.][]{andrews2013,pm2013} that displays less redshift-evolution than the mass-metallicity relation; this has lead some to suggest N/O as a fundamental tracer of galaxy evolution \citep[e.g.][]{pm2013,masters2016}. This result does appear to be calibrator-dependent, however, with \citet{hp2022} finding a residual inverse SFR--N/O trend when the [N\textsc{ii}]$_{6585}$$/$[O~\textsc{ii}]$_{3737, 3729}$ ratio is used to estimate N/O. This detail is important for interpreting the FMR. The lack of an equivalent N/O relation to the FMR would imply the FMR to be driven by short-term processes, with gas mass serving as a more fundamental parameter than SFR \citep[e.g.][]{bothwell2013,bothwell2016,brown2018}. If an equivalent N/O relation does exist, meanwhile, then galaxies' wider SFHs will be relevant for understanding the FMR \citep{hp2022}.}

On $\sim$kpc scales, an increasingly complex picture emerges for both abundance ratios. Local gas metallicities trend simultaneously with (local) stellar mass surface density (hereafter $\Sigma_*$) and (global) stellar mass \citep{sanchez2013,bb2016,gao2018}, with further residual metallicity trends apparent with local SFR along with related local parameters \citep[e.g.][]{hwang2019,boardman2022,baker2023}. Similar trends are found for local N/O abundances \citep{schaefer2022}. Regions with anomalously-low gas metallicities \citep[on the basis of $\Sigma_*$ and $\mathrm{M_\star}$;][]{hwang2019} also display elevated N/O for their O/H, which possibly points to recent metal-poor gas accretion in affected regions \citep{luo2021}. Star-forming galaxies generally possess negative radial gradients in both O/H and N/O \citep[e.g.][]{sanchez2014,belfiore2017}, with O/H gradients largely reproduced by local metallicity relations such as the density-metallicity relation within individual galaxies \citep{bb2016,boardman2022}. \citet{boardman2023} find galactocentric radius to frequently be an even stronger predictor of metallicity within individual MaNGA galaxies than density; this potentially reflects other radial trends in galaxies such as in gas-to-stellar mass fraction \citep[rises with increasing radius; e.g.][]{carton2015}, star-formation efficiency \citep[falls with radius; e.g.][]{leroy2008} and escape velocity \citep[falls with radius; e.g.][]{bb2018}.

Overall, gaseous abundances -- as parametrized through O/H and N/O -- appear to be tightly connected to \rev{SFHs}. Thus, we can learn more by considering the stellar populations \textit{in addition to} gaseous abundances, which is an approach that is becoming increasingly common. Studies such as \citet{lian2018} and \citet{frasermckelvie2022} simultaneously consider galaxies' stellar and gas-phase metallicities, finding stellar metallicity to be typically lower than gas metallicity when the two are on an equivalent base. Stellar population fitting also allows one to fit detailed SFHs to observed galaxy spectra, which then have the potential to be compared to gas-phase properties \citep[e.g.][]{greener2022}. It should be noted that stellar metallicity measurements generally make use of the iron-to-hydrogen ratio $\mathrm{[Fe/H]}$, with iron enrichment occuring over significantly longer timescales than for $\alpha$ elements such as oxygen \citep[e.g.][]{matteucci2001,matteucci2009,kobayashi2009}. The stellar $\mathrm{\alpha}$ abundance ratio $\mathrm{[\alpha/Fe]}$ is therefore conceptually similar to $\mathrm{\log(N/O)}$, in that both ratios are sensitive to complementary star-formation processes that operate on different timescales.  The past level of star-formation in a galaxy is thought to affect both sets of ratios significantly, with higher star-formation efficiencies being associated with greater early enrichment and hence with a higher gaseous (or stellar) metallicity at a given N/O (or $\mathrm{[\alpha/Fe]}$) \citep[e.g.][]{vincenzo2016,nidever2020}. Stellar $\mathrm{[\alpha/Fe]}$ studies have typically been restricted to older quiescent galaxies due to the difficulty of constraining $\mathrm{[\alpha/Fe]}$ in young ($<1$ Gyr) populations \citep[e.g.][]{conroy2018}, leading \citet{sanchez2021a} to propose using gaseous oxygen abundances as a stellar $\alpha$ proxy in star-forming galaxies. Efforts have also been made to measure star-forming galaxies' $\mathrm{[\alpha/Fe]}$ through absorption line indices, either through direct modelling with single stellar populations \citep[e.g.][]{scott2017} or through measurements of the offset from the Milky Way abundance pattern \citep{gallazzi2021}.

Full spectral fitting can provide detailed SFH constraints, and careful application of such fits on IFU data has yielded much insight into galaxy evolution on local scales \citep[e.g.][]{ibarramedel2016,goddard2017,gonzalezdelgado2017,lopezfernandez2018,ibarramedel2019,cf2022}. However, the results of full spectral fitting are highly model-dependent; furthermore, it is difficult to succinctly visualise the results of SFH fits for two-dimensional IFU data, in which there can be many fitted SFHs from spectra of a single galaxy and in which there can be $\sim$1000 galaxies or more in a considered sample. It is therefore desirable to also employ spectral indices, to provide a first-order characterisation of the SFH which lacks any dependence on models. In particular, the principal component analysis (PCA) method of \citet{wild2007} allows one to gain insight into the star-formation history of a spectrum from its Balmer absorption strength and its D4000 index, while being applicable to noisier spectra than traditional spectral indices. This method has various applications, such as studying SFHs of AGN host galaxies \citep{wild2007} and identification of post-starbursts \citep{wild2009}. The method was applied to MaNGA data by \citet{rowlands2018}, allowing for easy two-dimensional visualisation and analysis of spatially-resolved SFHs. 

\citet{rowlands2018} considered the gas-phase metallicities of star-forming regions in their galaxies, with the regions grouped according to their SFH classification. They found starburst regions to possess the lowest gas metallicities at most tested galactocentric radii, with post-starburst regions typically possessing metallicities between those of starburst regions and ordinary star-forming regions. Such results are consistent with the diluted gas metallicity profiles seen in highly star-forming galaxies \citep{ellison2018} along with the local anticorrelations between metallicity and SFR seen in many MaNGA galaxies \citep{sa2019}. \rev{In this paper we explore whether further insight can be gained by} also considering N/O in terms of spatially resolved SFHs, in addition to considering O/H as has been done previously. 

\rev{Here,} we employ the method of \citet{rowlands2018} to compare SFHs of star-forming MaNGA galaxy regions to their gaseous abundances, considering both gas-phase metallicity ($\mathrm{12 + \log(O/H)}$) and nitrogen-to-oxygen abundance ratio ($\mathrm{\log(N/O)}$) \rev{for the first time}. We probe the connection between SFH and gaseous abundance, considering in particular where spaxels of different SFH types fall upon the N/O--O/H plane. 

The layout of this paper is as follows. We present our sample and discuss relevant methods in \autoref{sampledata}, and we present our results in \autoref{results}. We discuss our findings in \autoref{disc}, and we summarise and conclude in \autoref{conclusion}. Throughout this work, we assume a Kroupa IMF \citep{kroupa2001} for the purpose of displaying stellar masses and densities, and we adopt the following standard $\Lambda$ Cold Dark Matter cosmology: $\mathrm{H_0} = 71$ km/s/Mpc, $\mathrm{\Omega_M}$ = 0.27, $\mathrm{\Omega_\Lambda}$ = 0.73.

\section{Sample \& data}\label{sampledata}

\subsection{Data acquisition}

We employ integral field spectroscopy from the SDSS-IV MaNGA survey. MaNGA observations were performed using the BOSS spectrographs \citep{smee2013} on the 2.5 meter telescope at Apache point observatory \citep{gunn2006}. MaNGA used a series of IFUs, with IFUs consisting of between 19 and 127 opical fibres of 2$^{\prime\prime}$ diameters \citep{law2016}. These IFUs employ hexagonal configurations, with three-point dithering employed to fully sample the field of view (FOV) \citep{drory2015,law2015}. MaNGA observed targets such that the logarithmic mass distribution was approximately flat, with redshifts ranging from roughly 0.01 to 0.15 \citep{yan2016b,wake2017}. MaNGA observations were reduced by the Data Reduction Pipeline \citep[DRP;][]{law2016,yan2016a}, which outputs $0.5^{\prime\prime} \times 0.5^{\prime\prime}$ spaxel datacubes. The MaNGA datacubes have a median point spread function (PSF) full-width at half-maximum (FWHM) of roughly 2.5$^{\prime\prime}$ \citep{law2016}, while the spectra have a resolution of $R \simeq 2000$ and cover a wavelength range of 3600--10000\AA. The Data Analysis Pipeline \citep[DAP;][]{belfiore2019,westfall2019} then computed various quantities relating to galaxies' stellar and gaseous components. MaNGA data and data products can be accessed via the SDSS science archive server\footnote{\url{https://data.sdss.org/sas/}} and also through the Marvin\footnote{\url{https://www.sdss.org/dr17/manga/marvin/}} interface \citep{cherinka2019}. The full MaNGA sample has been released as of SDSS DR17 \citep{sdssdr17} and consists of approximately 10,000 galaxies. 

We selected an initial parent sample by first considering all MaNGA galaxies in the Primary and Color-enhanced samples (hereafter referred to jointly as the ``Primary+ sample") along with the Secondary sample, which cover FOVs of approximately 1.5 and 2.5 half-light radii ($R_e$) respectively. A handful of galaxies were observed on multiple IFUs, resulting in duplicate observations; in such cases, we choose the observation with the highest combined signal-to-noise in the red and blue cameras \citep[see][for further information]{yan2016b}. We obtained for the resulting sample a number of parameters from the NASA-Sloan-Atlas \citep[NSA;][]{blanton2011} catalog: elliptical Petrosian stellar masses $\mathrm{M_\star}$, elliptical Petrosian half-light radii $\mathrm{R_e}$ and elliptical Petrosian axis ratios $b/a$. We restricted to galaxies with $b/a > 0.6$, in order to avoid edge-on galaxies. We also removed galaxies with quality flags `SEVEREBT', `UNUSUAL' or `CRITICAL', to avoid analysing galaxies with severe data quality problems. This yielded 6885 galaxies. 

We next proceeded to obtain mapped quantities for the parent sample from three sources: Pipe3D \citep{sanchez2016,sanchez2016b,sanchez2018, sanchez2022}, the MaNGA DAP, and SFH maps produced by applying the \rev{PCA} methods described in \citet{rowlands2018} on the DR17 MaNGA sample. We removed a further 18 galaxies that were not included in the MaNGA pipe3d summary table, and we also removed two galaxies which lacked maps from the DAP or PCA analysis. This resulted in a final parent sample of 6865 galaxies.

We obtained the following mapped stellar population properties from the Pipe3D maps: surface mass density $\Sigma_*$, the D4000 index and the $\mathrm{H\delta_A}$ index \citep{worthey1997}. The Pipe3D analysis is performed in MaNGA datacubes that are binned to achieve a signal-to-noise ratio of 50 across a given cube's field of view. Pipe3D computes $\Sigma_*$ from the stellar continuum, with a dust correction applied using a \citet{cardelli1989} extinction law with $R_V = 3.1$, and we multiply the calculated value by b/a to correct for the effects of inclination. We also applied galaxies' dezonification maps, \rev{which indicate the V-band flux contribution of spaxels to a given Pipe3d spectral bin}, to approximately correct for the effects of binning on $\Sigma_*$ values. Pipe3D calculates D4000 as the ratio between the average flux densities at 3750--3950 \AA\ and 4050--4250 \AA\, following \citet{bruzual1983}, while correcting the parameter for reddening as in Equation 2 of \citet{gorgas1999}; we obtain this parameter without performing any further corrections or adjustments. Broadly speaking, higher D4000 values correspond to older and more evolved stellar regions: the index positively correlates with stellar age \citep[e.g.][]{sanchez2016b} and negatively correlates with specific star-formation rate \citep[e.g.][]{brinchmann2004}, with higher stellar metal contents also yielding higher D4000 values \citep{gallazzi2005}. $\mathrm{H\delta_A}$ peaks when a population is dominated by A-type stars, which is associated with post-starburst-type histories \citep[e.g.][]{goto2003}.

From the DAP, we obtained measurements of the following emission lines: H$\alpha$, H$\beta$, [O~\textsc{iii}]$_{5008}$,[N\textsc{ii}]$_{6585}$, [S\textsc{ii}]$_{6718}$, [S\textsc{ii}]$_{6733}$ and [O~\textsc{ii}]$_{3737, 3729}$, along with the associated measurement errors. We corrected the fluxes for reddening by assuming an intrinsic Balmer decrement of 2.86 (valid for case B recombination, with $\mathrm{n_e = 100 \ cm^{-3}}$ and $\mathrm{T_e = 10000 \ K}$) \rev{and a \citet{fitzpatrick2019} reddening curve} with $R_V$ = 3.1. For spaxel spectra in which the measured Balmer decrement is less than 2.86, no dust correction is performed. We also obtained the H$\alpha$ equivalent width ($\mathrm{EW_{H\alpha}}$) from this dataset. We obtained from the PCA maps the values of the first two principal components (hereafter PC1 and PC2), the corresponding SFH classification, and the signal-to-noise around the D4000 region (hereafter $\mathrm{SNR_{D4000}}$). For any given spaxel, the SFH classification depends on the spaxel's PC1 and PC2 value along with the stellar mass of that spaxel's galaxy. The PC1 of a spectrum correlates substantially with its D4000 index; PC2 is associated with excess Balmer absorption, and so correlates with $\mathrm{H\delta_A}$ at a given PC1. The SFH classifications use the following labels: star-forming (hereafter SF), starburst (hereafter SB), post-starburst (hereafter PSB), green valley (hereafter GV) and quiescent. A small number of spaxels are assigned \rev{`unclassified'} status, with such cases mostly containing broad-line active galactic nuclei (AGN) spectra. 

\subsection{Selection of star-forming spaxels}\label{sfselection}

We calculate gas-phase metallicity 12 + log(O/H) along with nitrogen-to-oxygen abundance ratio N/O using strong emission line calibrators, as is discussed further in the following subsection. Strong line calibrators are generally determined from observations of star-forming H\textsc{ii} regions or else from photoionisation models, and most consequently are valid only for star-forming galaxy regions \citep[though see][for a potential solution]{kumari2019}. Thus, we needed to select star-forming spaxels from galaxies in our sample while ensuring that the quality of data in those spaxels was sufficiently high. To ensure good data quality, we require $S/N > 3$ for all relevant emission lines and we also require $\mathrm{SNR_{D4000}} > 4$. We use the [N\textsc{ii}] and [S\textsc{ii}] versions of the BPT diagram \citep{bpt} to selection star-forming regions, employing the \citet{kauffmann2003} demarcation line and the \citet{kewley2006} demarcation line respectively, while also requiring $\mathrm{EW_{H\alpha} > 14}$ \AA\ to minimise the impact of diffuse ionised gas \citep[DIG;][]{lacerda2018,valeasari2019}. We also remove spaxels with \rev{quiescent or unclassified SFHs} at this stage, along with applying loose quality cuts to $\Sigma_*$ and D4000 ($\Sigma_* > 1 \mathrm{M_\odot/pc^2}$, $0 < \mathrm{D4000} < 10$) and removing spaxels with metallicities above the range of our chosen calibrator (see next subsection for further details).

Motivated by the oversampling in MaNGA datacubes, and similarly to \citet{hwang2019}, we require galaxies to possess a minimum of twenty selected spaxels; we completely remove all spaxels belonging to galaxies below this threshold from our analysis. Our resulting galaxy sample consists of 2671 individual galaxies with \rev{1192515} star-forming spaxels. We note at this point that many of the described cuts only mildly affect the final sample size: if we were only to select spaxels from their BPT diagnostics and by requiring $\mathrm{EW_{H\alpha}} > 14$ with the same 20-spaxel threshold then applied for galaxies, we would arrive at a final sample size of 2675 galaxies and \rev{1200159} spaxels.

\begin{figure}
\begin{center}
	\includegraphics[trim = 1cm 10.8cm 1cm 9.5cm,scale=0.8,clip]{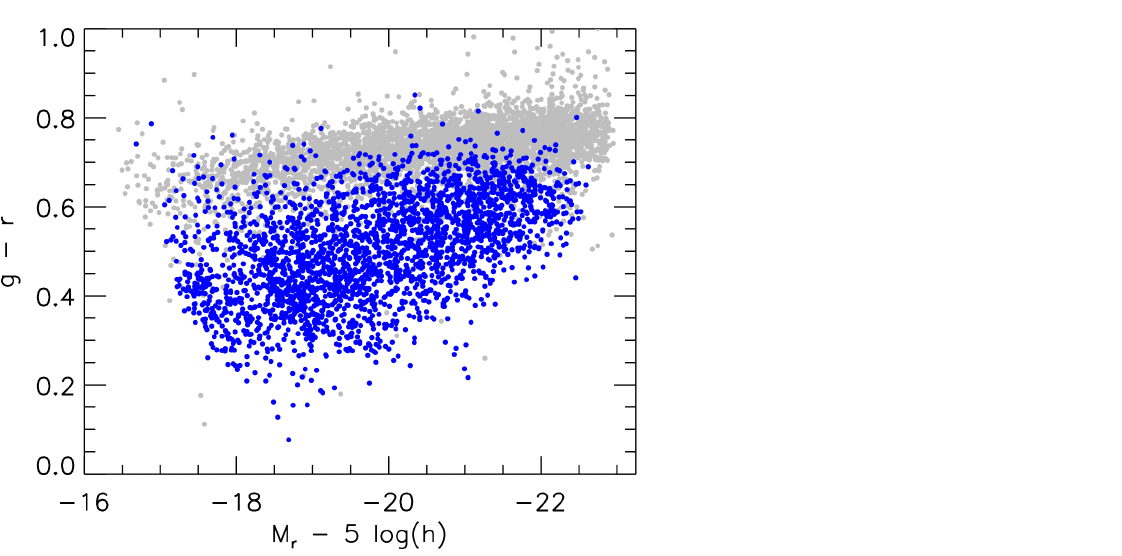} 
	\caption{Final galaxy sample (blue points) and remaining parent sample galaxies (grey points), presented in terms of $g-r$ colors and $r$-band absolute magnitudes. The final sample is dominated by blue galaxies and tends towards slightly fainter magnitudes than the parent sample, due to our implicit selection on star-formation status.}
	\label{colormag}
	\end{center}
\end{figure}

In \autoref{colormag}, we show our parent and final galaxy samples in terms of $g-r$ colors and $r$-band absolute magnitudes, using elliptical Petrosian magnitudes from the NSA catalog. Our sample selection strategy implicitly requires galaxies to be experiencing ongoing star-formation, resulting in the final sample tending towards bluer colors and slightly fainter magnitudes than the parent sample. We also note that all sample spaxels are star-forming on the basis of emission line metrics, irrespective of their SFH label.

\subsection{Gas-phase abundance calculation}

A range of gas abundance calibrators have been proposed in the literature, in which gas metallicities -- along with N/O abundance ratios -- are estimated from emission line flux ratios. Many calibrators have been derived from photoionization models, in which input chemical abundances allow predicted emission line ratios to be recovered \citep[e.g.][]{blanc2015,Dopita_2016_EmLineDiagnostic,morisset2016}. However, such calibrators depend on major assumptions concerning the physical behaviours of ionizing populations' atmospheres; moreover, such calibrators assume O/H to trend with ionisation strength or else with abundance ratios such as N/O, implicitly or otherwise \citep[for a summary, see Section 3.1 of][and references therein]{sanchez2017a}. 

Observations of star-forming H\textsc{ii} regions provide an alternative to photoionisation models. Gas-phase metallicities can be determined from ratios of auroral to nebular emission lines in the so-called `direct method', with N/O then also possible to estimate by taking the ratio of ionised nitrogen to oxygen ($N^{+}/O^{+}$) as a proxy \citep[e.g.][]{marino2013,la2020}, which allows relations between these and strong emission line ratios to then be determined. This method has traditionally been problematic at super-solar metallicities even for nearby H\textsc{ii} regions, for which auroral lines become increasingly weak. \citet{curti2017,curti2020} address this problem via stacking of high-metallicity galaxy spectra from SDSS, allowing a variety of calibrators to be determined over a wide metallicity range; these determinations are carried out via the direct method, with emission fluxes estimated using expected ratios \citep{pilyugin2005,pilyugin2006} where necessary. 

Our calculation of gas metallicity and N/O largely follows the methodology of \citet{luo2021} and references therein. We use the RS32 calibrator of \citet{curti2020}, where $\mathrm{RS32 = [O~\textsc{iii}]_{5007}/H\beta + [S\textsc{ii}]_{6717,6731}/H\alpha}$, to derive the gas metallicity. We use the N2O2 relation of \citet{la2020} to calculate N/O, in which $\mathrm{N2O2 = [N\textsc{ii}]_{6585}/[O~\textsc{ii}]_{3727, 3729}}$. These calibrators are chosen for their independence to one another and for their relative insensitivity to the ionisation parameter. It should be noted in particular that the RS32 calibrator does not depend on the behaviour of nitrogen lines, making it resistant to the biases relating to N/O variations reported for certain other calibrators \citep{pm2009,schaefer2020}. The RS32 calibrator is double-valued, peaking and turning over at $\mathrm{12 + \log(O/H)} \simeq 8$; given that massive galaxies typically possess gas metallicities well in excess of this, we assume all spaxels to be on the upper metallicity branch in our analysis. The calibrator is fitted to metallicities up to $\mathrm{12 + \log(O/H)} = 8.85$ in \citet{curti2020}, and so we remove the small number of spaxels at this metallicity or higher from our analysis; \rev{we also remove spaxels for which RS32 is measured to be higher than than the maximum fitted value in \citet{curti2020}}. 

\subsection{Grouping by SFH and control sample selection}

At this point, we have selected a sample of MaNGA spaxels (and their associated galaxies), for which we have a number of computed quantities related to their stellar and gaseous components. Amongst these quantities are the first two principal components from PCA fitting along with the corresponding SFH classifications. The PC1--PC2 classification boundaries are updated from \citet{rowlands2018}, to improve agreement with \citet{chen2019} selections \citep[][\rev{their appendix A}]{otter2022} and motivated further by improvements to MaNGA spectrophotometric calibration, and we show our applied boundaries in \autoref{boundaryfig}. \rev{Further information on the boundaries can be found in Appendix A.}

\begin{figure}
\begin{center}
	\includegraphics[trim = 1cm 0cm 1cm 0cm,scale=0.6]{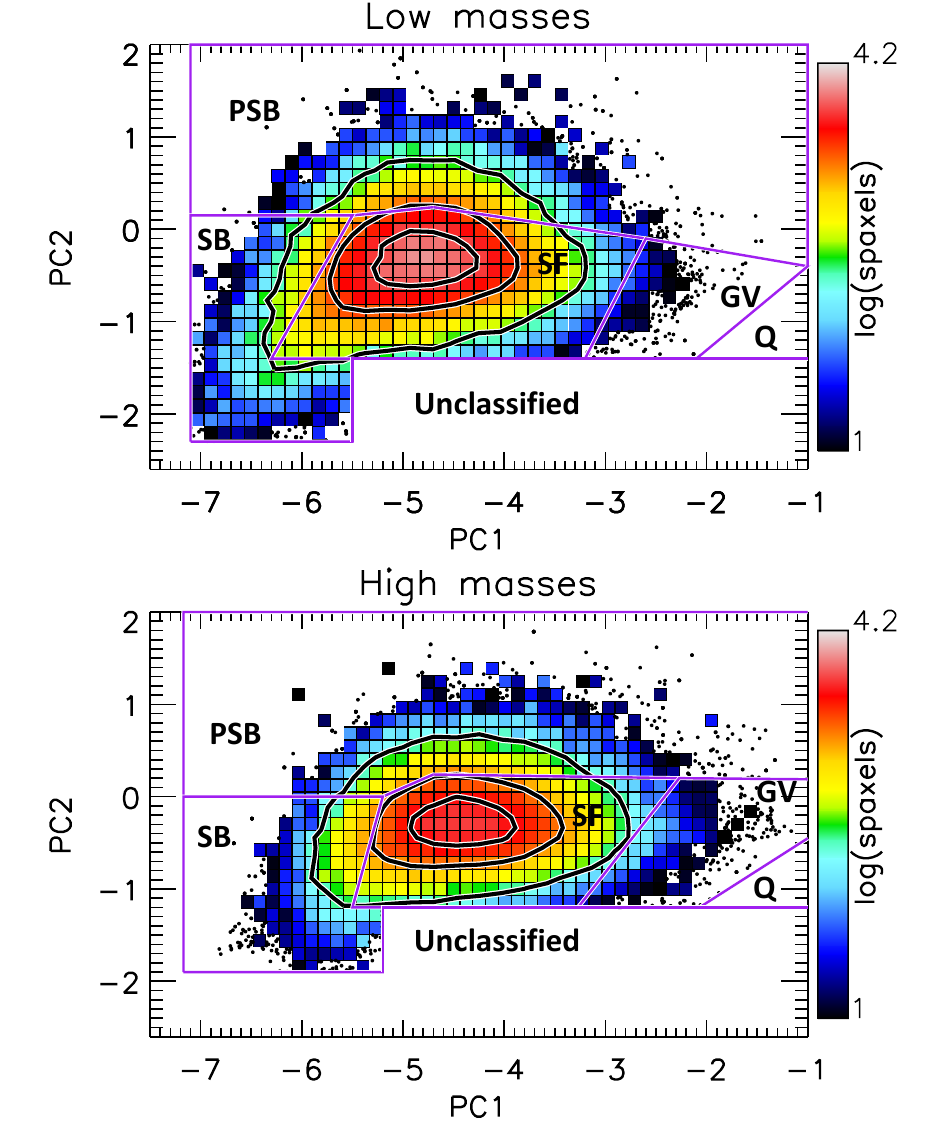} 
	\caption{PCA classification boundaries with our final spaxel sample plotted as a color map. Each square represents a bin of at least 10 spaxels, and spaxels outside of these bins are shown as black dots. The boundaries employ a dividing NSA elliptical Petrosian stellar mass of $10^{10} \mathrm{M_\odot}$, with the default NSA cosmology ($h = 1, \Omega_M = 0.3, \Omega_\Lambda = 0.7$) used in this specific case. Using the shown boundaries, we classify spaxels as star-forming (SF), starburst (SB), post-starburst (PSB), green valley (GV), quiescent (Q) or \rev{unclassified}. By design, our final spaxel sample includes no \rev{quiescent or unclassified spaxels}. The contours encompass $\sim$95\%, $\sim$65\% and $\sim$30\% of the spaxels.}
	\label{boundaryfig}
	\end{center}
\end{figure}

We now proceed to group the spaxel sample according to their SFH label, resulting in separate subsamples of SF spaxels, SB spaxels, PSB spaxels and GV spaxels. The sizes of these subsamples are as follows:

\begin{itemize}
\item SF subsample: \rev{1013826}
\item SB subsample: \rev{64980}
\item PSB subsample: \rev{110372}
\item GV subsample: \rev{3337}
\end{itemize}

We reiterate that \textit{all} of these subsamples are `star-forming' in terms of their ionised gas diagnostics, as was described in Section 2.2; the names of these subsamples, therefore, specifically refer to their stellar continuum-based classifications. This is an important caveat for the PSB and GV subsamples, which are not necessarily representative of PSB and GV regions (as determined from PCA) as a whole. Indeed, certain other PSB selection criteria \citep[e.g.][]{goto2007,chen2019} would reject our PSB subsample altogether, due to such criteria requiring little to no $\mathrm{H\alpha}$ emission. The PCA approach allows us to detect and select PSB regions with declining-but-present star-formation, which is not possible when applying strict cuts on emission lines. 

To facilitate closer comparisons between the SF spaxels and the others, we construct control samples for each of the non-SF spaxel subsamples. For each spaxel in the non-SF subsamples, we select a corresponding spaxel from the SF subsample; this SF spaxel is selected so as to belong to a different galaxy while minimising $\Delta(\mathrm{M_\star})^2 + \Delta(\mathrm{R/R_e})^2$, where $\Delta(\mathrm{M_\star})$ describes the logarithmic difference in $\mathrm{M_\star}$ and $\Delta(\mathrm{R/R_e})$ describes the linear difference in $\mathrm{R/R_e}$ between a given SF spaxel and the non-SF spaxel being considered. We refer to these control samples as the `SB control', `PSB control' and `GV control' subsamples for the remainder of this article. Our use of $\mathrm{M_\star}$ and $\mathrm{R/R_e}$ is motivated by the strong correlation between local gas metallicities and galaxy stellar masses \citep[e.g.][]{gao2018}, along with the observed tight trends between metallicity and galactocentric radius in individual galaxies \citep[e.g.][]{boardman2023}. We note that we obtain similar results if we instead use $\log(\Sigma_*)$ as the second selection parameter instead of $\mathrm{R/R_e}$; such a selection would be well-motivated by past reports of strong density-metallicity relations \citep[e.g.][]{bb2016}, but it would also be more model-dependent.

\section{Results}\label{results}

\begin{figure}
\begin{center}
	\includegraphics[trim = 1.5cm 9cm 0cm 11.5cm,scale=0.85]{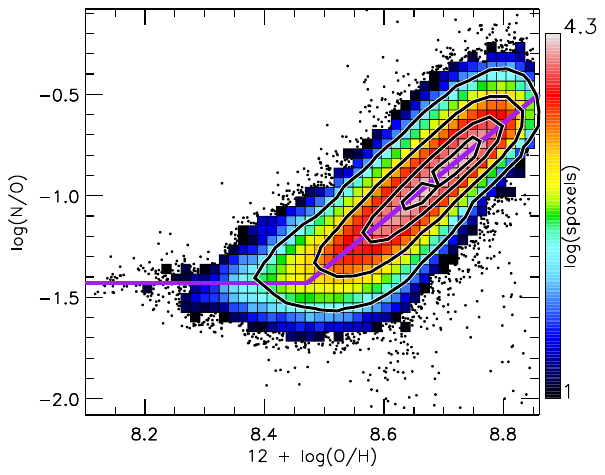} 
	\caption{Contours of $\log(\mathrm{N/O})$ plotted against gas metallicity for the star-forming spaxel subsample, along with the fitted N/O--O/H relation (purple lines) \rev{and the distribution of the full sample (coloured squares and dots). Each square represents a bin of at least 10 spaxels, with dots representing individual spaxels outside of these bins.} The contours encompass $\sim$99\%, $\sim$90\%, $\sim$60\% and $\sim$30\% of the \rev{SF} spaxels. \rev{A further 20 spaxels with anomalously low $\log(\mathrm{N/O})$ are not shown}.} 
	\label{novoh_sf}
	\end{center}
\end{figure}

In \autoref{novoh_sf}, we show the \rev{fitted} O/H--N/O relation for the SF subsample\rev{, along with showing our full spaxel sample in O/H--N/O space}. As expected from past works, we obtain a tight trend between the two indicators. We perform a least-absolute-deviation straight-line fit to parameterise this relation \rev{for the SF subsample}, while \rev{applying a fixed} $\log(\mathrm{N/O})$ plateau of $-1.43$ \citep{andrews2013} as in \citet{luo2021}. We first perform the straight-line fit on the full SF subsample, and then we iteratively perform the fit on spaxels for which the fitted N/O--O/H line is above the N/O plateau. We obtain from this \rev{process $\log(\mathrm{N/O}) = 2.397[12+\log(\mathrm{O/H})] - 21.732$ for metallicities above $12 + \log(\mathrm{O/H}) = 8.471$, with} $\log(\mathrm{N/O})$ set to $-1.43$ below this point. 

\rev{We note from \autoref{novoh_sf} a small number of spaxels with anomalously low N/O values, which typically possess highly elevated Balmer decrements with $E(B-V) > 1$. We ascribe such high Balmer decrements to measurement errors in the $H\beta$ emission flux. Given our substantial overall sample size, these anamolous spaxels have no bearing on our reported results.}

\begin{figure*}
\begin{center}
	\includegraphics[trim = 0.5cm 10.4cm 1cm 13cm,scale=1.1]{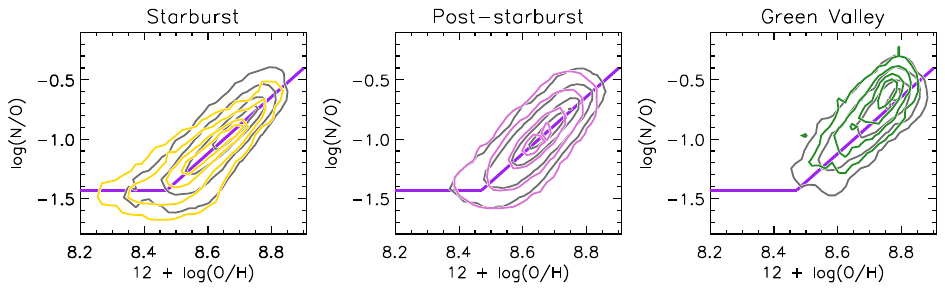} 
	\caption{Contours of $\log(\mathrm{N/O})$ plotted against gas metallicity for the starburst, post-starburst and green valley spaxel subsamples (coloured lines), along with the N/O--O/H relation fitted to the star-forming sample (purple lines) and N/O--O/H contours from the corresponding control samples (dark grey lines). The contours encompass $\sim$99\%, $\sim$90\%, $\sim$60\% and $\sim$30\% of spaxels in a given subsample.} 
	\label{novoh_nonsf}
	\end{center}
\end{figure*}

Next, we show in \autoref{novoh_nonsf} the remaining three spaxel subsamples in terms of N/O and O/H, with the corresponding control samples also shown. We find the SB subsample to preferentially possess lowered O/H and N/O abundances compared to the SB control sample; the PSB subsample behaves far more similarly to its control sample, with only a mild preference for lower abundance ratios at the high-metallicity end. Green valley spaxels, meanwhile, typically possess slightly elevated N/O ratios at a given metallicity. We find a tight N/O--O/H relation in all cases: the SB and and PSB subsamples appear to follow essentially the same relation as the SF subsample, with the GV subsample being offset to mildly higher N/O values.  

\begin{figure*}
\begin{center}
	\includegraphics[trim = 1.5cm 1.5cm 0cm 19cm,scale=0.95]{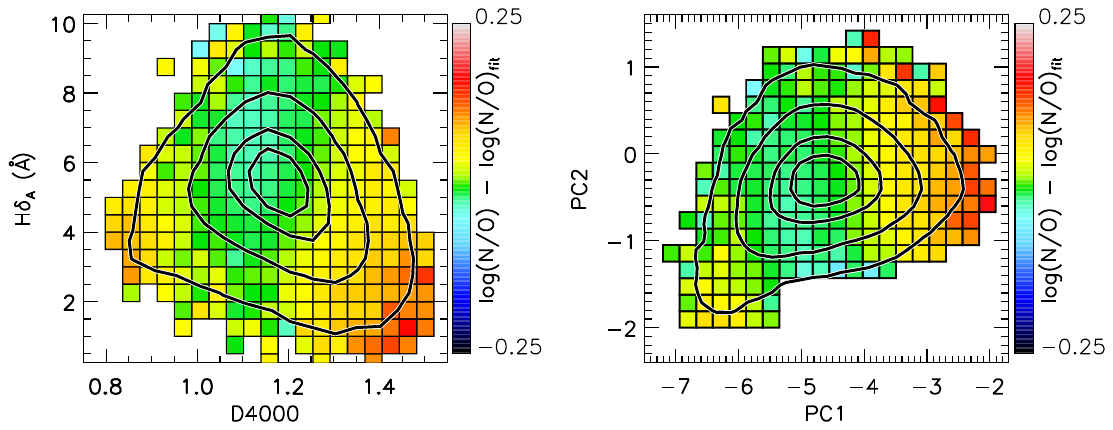} 
	\caption{Median offset of $\log(\mathrm{N/O})$ (for all spaxels) from the \rev{N/O--O/H relation fitted to} the star-forming spaxel subsample, in terms of D4000 and $\mathrm{H\delta _A}$ (left) and in terms of PC1 and PC2 (right). We find the median offset to trend mostly with D4000 (left) and with PC1 (right), with high D4000 and PC1 values associated with the green valley component of our spaxel sample. All displayed bins contain at least 50 spaxels. The black contours encompass $\sim$99\%, $\sim$90\%, $\sim$60\% and $\sim$30\% of spaxels in the full spaxel sample.} 
	\label{no_offsets}
	\end{center}
\end{figure*}

The PCA method essentially depends on the behaviour of the D4000 and Balmer absorption features; thus, it is informative to consider the behaviour of $\log(\mathrm{N/O})$ in terms of these parameters. In the left panel of \autoref{no_offsets} we plot the median difference between $\log(\mathrm{N/O})$ and the value predicted from the star-forming subsample's N/O--O/H relation, in bins of D4000 and $\mathrm{H\delta _A}$. From this figure, we show that high N/O offsets -- that is, N/O values above the best-fit line from star-forming spaxels -- are largely associated with high D4000 values, which corresponds to green-valley classifications. We also see significant median N/O offsets at the very low D4000 end, which corresponds to starburst SFHs.

In the right panel of \autoref{no_offsets} we plot the median N/O offset in terms of PC1 and PC2, from which see that the offset trends almost entirely with PC1. This follows naturally from the left panel: PC1 trends tightly with D4000, and so naturally trends tightly with the N/O offset, while PC2 traces the offset of spaxels from the D4000-$\mathrm{H\delta _A}$ relation. We also see notably raised N/O values at the very lowest PC2 and PC1 values, corresponding to the most extreme starburst regions of our spaxel sample. \rev{We note at this point that the precise adopted N/O plateau value has only a mild impact on the appearance of \autoref{no_offsets}, with effects largely confined to the starburst region of the parameter space. Higher plateau values make the N/O offsets around the starburst region less prominent, while lower plateau values make these N/O offsets more prominent, though a value of -1.43 appears well-motivated both from previous work \citep{andrews2013} and from our own measurements (\autoref{novoh_sf}).}

\begin{figure*}
\begin{center}
	\includegraphics[trim = 1.5cm 1.5cm 0cm 19cm,scale=0.95]{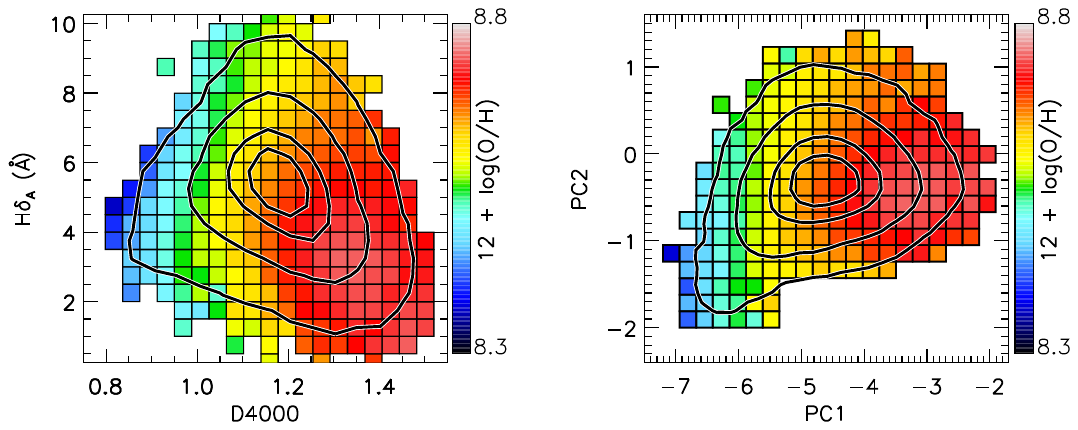} 
	\caption{Median gas metallicity in terms of D4000 and $\mathrm{H\delta _A}$ (left) and in terms of PC1 and PC2 (right). We find metallicity to trend chiefly with PC1 and D4000, but we also find metallicity to decrease somewhat at high $H\delta _A$ values. All displayed bins contain at least 50 spaxels. The black contours encompass $\sim$99\%, $\sim$90\%, $\sim$60\% and $\sim$30\% of spaxels in the full spaxel sample.} 
	\label{oh-merged}
	\end{center}
\end{figure*}

In \autoref{oh-merged}, we plot the median gas-phase metallicity in the D4000-$\mathrm{H\delta _A}$ and PC1--PC2 parameter spaces. We note a significant trend between metallicity and D4000 (and by extension betwen metallicity and PC1), consistent with previous work \citep[e.g.][]{sm2020,boardman2023}. We find however that  high $\mathrm{H\delta _A}$ values are associated with somewhat lower median metallicites at a given D4000, which is also reflected in PC1--PC2 space.

\begin{figure*}
\begin{center}
	\includegraphics[trim = 2cm 3.3cm 0cm 1.5cm,scale=1.5,clip]{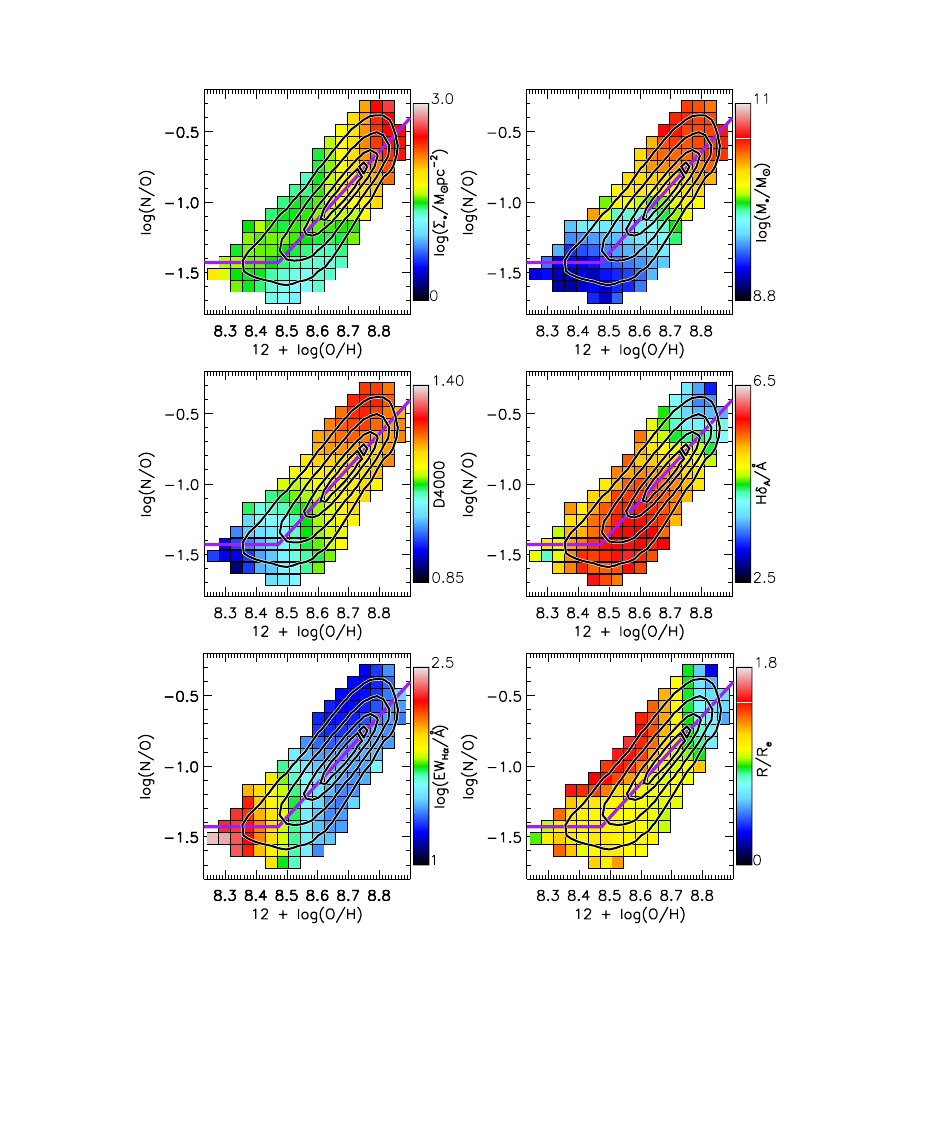} 
	\caption{Median parameter values of spaxels in bins of $\log(\mathrm{N/O})$ and $12+\log(\mathrm{O/H})$. We show results for stellar surface density (top left), the associated galaxy stellar mass (top right), the D4000 index (middle left), the $\mathrm{H\delta_A}$ index (middle right), the $\mathrm{H\alpha}$ equivalent width (bottom left) and galactocentric radius normalised to each galaxy's effective radius (bottom right). All displayed bins contain at least 50 spaxels. The black contours encompass $\sim$99\%, $\sim$90\%, $\sim$60\% and $\sim$30\% of spaxels in the full spaxel sample.} 
	\label{no-oh-maps}
	\end{center}
\end{figure*}

Next, we plot in \autoref{no-oh-maps} the behaviour of the full spaxel sample in N/O--O/H space in terms of various quantities: $\Sigma_*$, normalised galactocentric radius $\mathrm{R/R_e}$, D4000, $\mathrm{EW_{H\alpha}}$, $\mathrm{H\delta _A}$ and the $\mathrm{M_\star}$ of each spaxel's corresponding galaxy, with the first five parameters being local and the final one being global across a given galaxy's spaxels. As expected, we find higher metallicities to be associated with higher values of $\Sigma_*$, D4000 and $\mathrm{M_\star}$ along with lower values of $\mathrm{R/R_e}$ and $\mathrm{H\delta _A}$. We find elevated N/O values at low-to-intermediate metallicities to be associated with larger values of $\mathrm{R/R_e}$, similarly to what \citet{luo2021} report for anamolously-low-metallicity regions in MaNGA. We also find elevated N/O values at high metallicities to be associated with increased D4000 values and decreased $\mathrm{H\delta _A}$ values, which is unsurprising given the behaviour of the GV subsample discussed previously. We find $\mathrm{EW_{H\alpha}}$ to behave in an inverse manner to D4000 in the N/O--O/H plane: the highest values occur at low abundances, with higher-elevated N/O ratios at high metallicities being associated with the lowest $\mathrm{EW_{H\alpha}}$ values. Such inverse behaviour in unsurprising, since $\mathrm{EW_{H\alpha}}$ and D4000 are known to anti-correlate when star-formation is ongoing \citep[e.g.][and references therein]{cc2023}. 

Considering the Spearman rank correlation coefficient $\rho$, with find D4000 and $\mathrm{M_\star}$ to both correlate slightly stronger with N/O than with the gas metallicity. For D4000, we find $\rho = 0.55$ and $\rho = 0.52$ for N/O and for metallicity respectively; we find $P << 0.01$ in both cases and in all subsequent cases, where $P$ describes the probability of obtaining coefficients equal to or higher than what is measured in the case where there is no underlying correlation. For $\mathrm{M_\star}$, we find $\rho = 0.68$ and $\rho = 0.62$ for N/O and for metallicity respectively. In the case of $\Sigma_*$, we find $\rho$ = 0.60 and $\rho$ = 0.61 for N/O and for metallicity respectively, meaning what we obtain a slightly higher correlation with metallicity. As such, we find both gas abundance parameters to correlate strongly with properties relating to the stellar components, as has been reported repeatedly for the parameters individually in MaNGA and elsewhere \citep[e.g.][]{sanchez2013,bb2016,sanchez2017a,hwang2019,ep2022,schaefer2022}; however, neither parameter is clearly more or less informative than the other. We also note that excluding GV subsample spaxels has very little impact on these comparisons, due to the relatively small size of the GV subsample.

To summarise, we find that SB spaxels tend to have lower gas abundances than SF spaxels while following essentially the same N/O--O/H relation, with PSB spaxels behaving similarly to SF spaxels. We also find GV spaxels to have higher metallicities on average, while peaking at higher N/O values and slighly lower O/H values when compared to their star-forming control sample. We found complex relationships between O/H and N/O and other parameters relating to galaxy's stellar component: higher values of N/O and O/H are associated with higher values of $\mathrm{M_\star}$, $\Sigma_*$ and D4000 as expected along with lower $\mathrm{R/R_e}$, but elevated N/O values (for a given O/H) are found to be associated with higher $\mathrm{R/R_e}$ (at low-to-intermediate metallicty) or else with higher D4000 (at high metallicity).  

\section{Discussion}\label{disc}

In this work, we have employed MaNGA data to assess the connection between SFH and gas-phase abundance on local scales. We considered the behaviour of different spaxel samples in N/O--O/H space, with samples grouped by their classifications from the \citet{rowlands2018} PCA method. We also considered local stellar properties (D4000, $\Sigma_*$), along with $\mathrm{R/R_e}$ and $\mathrm{EW_{H\alpha}}$, as a combined function of N/O and O/H. We found that the highest N/O abundances are associated with high D4000 values and by extension with high PC1 values; we further found that such regions are offset from the usual N/O--O/H relation, possessing low metallicities for their N/O. Equivalently, we found the green valley spaxel sample to display elevated N/O when compared to its star-forming control. We found starburst spaxels to display reduced gas abundances compared to star-forming regions, while we found post-starburst regions to behave very similarly in N/O--O/H space to their star-forming control sample. 

\begin{figure}
\begin{center}
\includegraphics[trim = 0cm 10.4cm 1cm 11.5cm,scale=1]{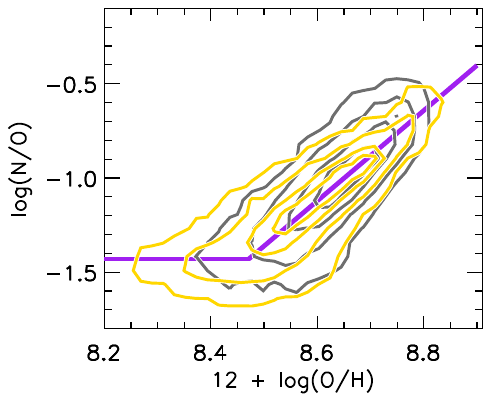} 
	\caption{Contours of $\log(\mathrm{N/O})$ plotted against gas metallicity for the SB spaxel subsample (gold), along with the fitted N/O--O/H relation (purple lines) and a control sample selected from PSB spaxels (grey contours). The contours encompass $\sim$99\%, $\sim$90\%, $\sim$60\% and $\sim$30\% of a given spaxel subsample.} 
	\label{novoh_sb_psbcontrol}
	\end{center}
\end{figure}

To further assess the behaviour of starburst and post-starburst spaxels, we compare in \autoref{novoh_sb_psbcontrol} the starburst sample to a new `control sample' selected from post-starburst samples. This control sample was selected using a combination of $\mathrm{M_\star}$ and $\mathrm{R/R_e}$, just as for the control samples already discussed. In this case, we find the starburst and `control' PSB sample to differ in two key ways: the metallicity and N/O of starbursts is lower on average than for PSB regions, while starburst spaxels' metallicity is somewhat elevated compared to PSB spaxels at the highest N/O values. Our findings therefore support a view in which gas-phase abundances typically increase rapidly after a starburst, such that post-starburst spaxels behave very similarly to star-forming spaxels in terms of their chemical abundances. A rapid enrichment of gas is similar to what \citet{leung2023} report for MaNGA post-starburst regions' stellar populations. \citet{leung2023} obtain the metallicities of old and young populations in these regions separately by performing full spectral fitting with the Bagpipes code \citep{carnall2018,carnall2019}, and they find younger populations -- those formed during the starburst -- to typically be more metal-rich than their older counterparts at a level reasonably consistent with the metallicity offset reported between star-forming and quiescent galaxy samples \citep[e.g.][]{peng2015}. It should be noted that a rise in metallicity over successive stellar generations is entirely expected, and it has been reported previously from fossil record analyses of spectra \citep[e.g.][]{panter2008}.

From our results, the picture that emerges is as follows: starburst regions are typically diluted significantly when a starburst event first occurs, with regions then re-enriching rapidly as they move into the post-starburst phase. Only the highest-abundance starburst regions behave differently, preferring slightly higher metallicities to post-starburst regions at a given N/O (\autoref{novoh_sb_psbcontrol}); this perhaps points to additional relevant physics in such cases. Nonetheless, the similarity of post-starburst to star-forming regions in N/O--O/H space implies that gas abundances are largely insensitive to the precise form of a local SFH at late times, with bursty SFHs ultimately producing similar results to the more gradual SFHs of ordinary star-forming regions.

Our results on the relative behaviour of star-forming and starburst regions are in good consistency with \citet{rowlands2018} along with other recent MaNGA work \citep[e.g.][]{ellison2018,sa2019}: high star-formation is associated with diluted gas abundances, similar to what was observed for gas metallicity on galaxy scales \citep[e.g.][]{ellison2008}. Our results are also consistent with merger simulations, in which infalling low-metallicity gas dilutes metals within galaxies' gas \citep{montuori2010,rupke2010,perez2011}. A picture therefore emerges in which infalling metal-poor gas dilutes the existing gas supply and triggers localised starburst events, with gas subsequently re-enriching rapidly.

The above-described picture for starburst regions is similar to what is argued for so-called anamously-low metallicity (ALM) regions in \citet{hwang2019}: they identify a significant population of spaxels with low metallicity for their associated $\Sigma_*$ and $M_\star$ values, and argue metal-poor gas inflows as the cause of such spaxels' offsets. The increasing fraction of starburst spaxels at large radii \citep{rowlands2018} is similar to what \citet{luo2021} report for ALM spaxels, further suggesting a connection between local starburst events and ALM regions. However, \citet{luo2021} find ALM spaxels to display elevated N/O ratios for their metallicity which they likewise ascribe to metal-poor inflows, which we do not find for starburst regions as a whole. Our results should not be viewed as being inconsistent with \citet{luo2021}, since starburst and ALM regions represent different samples. Gas metallicity will begin to rise quickly following a starburst due to the short ($\sim$10 Myr) timescales in which oxygen is produced, which will quickly move a region back towards the expected O/H-N/O relation. The average metallicity offset between starburst and starburst control spaxels is 0.070 dex, which is smaller than the 0.111 dex offset used in \citet{luo2021} to define ALM regions.  Thus, while some overlap can be expected between starburst and ALM spaxels, one should not expect them to behave identically. 

In terms of underlying physics, chemical evolution in galaxies can be broadly considered in terms of three physical processes: gas inflow, star-formation (and subsequent enrichment of the ISM), and outflow of metal-enriched gas \citep[e.g.][]{ferreras2000}. Metal-poor inflows will initially deplete O/H preferentially to N/O \citep[e.g.][]{koppen2005,belfiore2015}, resulting in anomalously low metallicities along with elevated N/O ratios \citep{luo2021}. However, we find that starburst regions in general are \textit{not} associated with N/O elevations: from the right panel of \autoref{no_offsets}, only the youngest starbursts -- those with the smallest PC2 values -- are associated with significant N/O elevations for their metallicity. Considering both the expected timescale for oxygen enrichment, and considering also the expected progression of an extreme starburst in PC1--PC2 space \citep{pawlik2016}, this suggests oxygen enrichment effectively `enriches away' the N/O enhancement in just a few tens of Myr after a starburst event. An extreme starburst can then be expected to reach the post-starburst phase within 500 Myr \citep{pawlik2016}, with the dilution from a starburst having largely been eliminated by that time. A similar timescale for ALM regions can be arrived at through various means including a closed box evolution model and from considering the stellar ages of low-metallicity regions, as argued by \citet{hwang2019}, which further supports an overlap between starburst events and ALM regions. Such an overlap would be entirely expected if the two both result from metal-poor inflows, which would serve to produce ALM regions with low metallicities and elevated N/O regions; a localised starburst could then be triggered in a subset of cases, leading to rapid oxygen enrichment and to an evolution away from ALM-like behaviour.

It is important to consider that infalling gas need not be pristine or from outside a galaxy. Enriched gas can be redistributed from a galaxy's centre to its outer parts via wind recycling \citep{oppenheimer2008}, which provides a possible explanation for the relatively high metallicities of massive galaxies' outer discs \citep[e.g.][]{bresolin2012,belfiore2017} and which can also explain N/O abundances that are elevated for the metallicity \citep{belfiore2015}. Radial gas flows \citep[e.g.][]{trapp2022,wang2022} could also play a role in supplying gas for a localised starburst event. Both wind recycling and gas flows provide possible explanations for the high metallicity end of the starburst sample. However, we disfavor these explanations for the majority of starburst spaxels, due to these spaxels having significantly lower average abundances than their star-forming controls. We also note that significant inward radial flows have a limited observational basis, with studies typically not detecting significant radial flows in galaxies \citep{wong2004,trachternach2008,teodoro2021} aside from in \citet{schmidt2016}. Finally, the possibility exists that starburst regions possess higher metallicity gas that is obscured by significant dust content, as has been reported within ultraluminous infrared galaxies in the nearby Universe \citep{chartab2022}.

We also noted a significant population of high-metallicity spaxels with elevated N/O ratios for their metallicity and with relatively high D4000 values.  The GV subsample is dominated by such spaxels, as can be seen in \autoref{novoh_nonsf}. However, the GV subsample is comparatively small and is \textit{not} the dominant contributor to this population, such that the appearance of \autoref{no-oh-maps} is essentially unchanged if the GV subsample is excluded. These spaxels are typically found in high-mass galaxies and away from those galaxies' centres, consistent with previous MaNGA work \citep{belfiore2017,schaefer2020}, though they are typically found at lower values of $\mathrm{R/R_e}$ compared to high-N/O spaxels of lower metallicity. We also observe similar behaviour if we use unbinned maps of $\mathrm{D_n4000}$ from the DAP in place of the Pipe3D D4000 values.

Our applied strong-line calibrators for O/H and N/O are both comparatively insensitive to the ionization parameter, as can for instance be seen in \citet[][their Figures 3 and 6]{dopita2000}; as such, ionization parameter variations are not a likely factor in our results for high-D4000 regions. DIG poses a far more significant concern, as it is known to impact upon both gas metallicity measurement and emission spectra interpretation more generally \citep[e.g.][]{metha2022}. DIG emission is powered by different sources than H\textsc{ii} regions, and it is likely a product of multiple individual processes \citep[e.g.][]{belfiore2022}. MaNGA's relatively course spatial sampling means that one cannot truly isolate `pure' star-forming H\textsc{ii} regions, which in turn means that spaxels termed `star-forming' will still contain some amount of diffuse emission \citep{valeasari2019}; a similar caveat exists for the CALIFA survey \citep{lacerda2018}, albeit to a lesser extent than in MaNGA due to CALIFA's smaller spatial scales. Controlling for metallicity, an analysis of MaNGA data suggests that DIG will serve to enhance the RS32 indicator while leaving N2O2 largely unaffected \citep{zhang2017}. Such an RS32 enhancement would spuriously create exactly the effect we see in our high-D4000 spaxels.   

\begin{figure}
\begin{center}
	\includegraphics[trim = 1.5cm 8cm 1cm 13.5cm,scale=1.05]{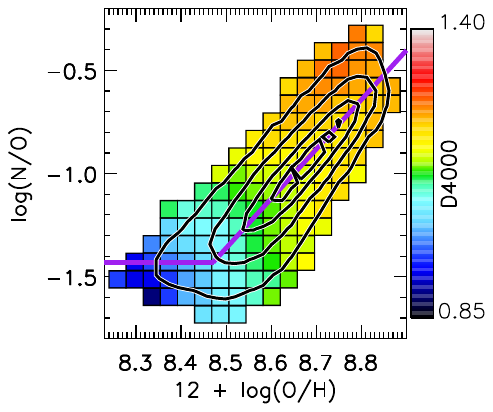} 
	\caption{Median D4000 values of spaxels in bins of $\log(\mathrm{N/O})$ and $12+\log(\mathrm{O/H})$, with the sample $\mathrm{EW_{H\alpha}}$ cut raised to $\mathrm{EW_{H\alpha}} > 20$. All displayed bins contain at least 50 spaxels. The black contours encompass $\sim$99\%, $\sim$90\%, $\sim$60\% and $\sim$30\% of spaxels in the full spaxel sample.} 
	\label{no-oh-d4}
	\end{center}
\end{figure}

To further reduce the impact of DIG, we tried an even stricter  $\mathrm{EW_{H\alpha}}$ cut of $\mathrm{EW_{H\alpha}} > 20$ and then again considered D4000 as a function of O/H and N/O. Applying all other cuts as described in Section \ref{sfselection} we obtain a sample of \rev{2476} galaxies and \rev{937661} spaxels. We show in \autoref{no-oh-d4} that we continue to see similar D4000 behaviour in this case. As such, we believe that potential physical explanations should also be explored.

As has been previously reported \citep[e.g.][]{belfiore2017,schaefer2022}, N/O correlates significantly with stellar mass. \citet{belfiore2017} provide two potential explanations for this: variations in star-formation efficiency \citep[SFE; e.g.][]{molla2006,wang2021}, and galactic winds \citep{oppenheimer2008}. A lower SFE would reduce the number of early core-collapse supernovea and hence would reduce early oxygen enrichment, resulting in N/O beginning to rise at earlier metallicities and producing an offset from the usual N/O--O/H relation \citep{molla2006,vincenzo2016}. Galactic winds could transport some amount of high-metallicity high-N/O gas to a galaxy's metal-poor outer-parts, preferentially raising the N/O of those regions \citep[e.g.][]{belfiore2015}. SFE is known to decline with galactocentric radius in galaxies \citep{leroy2008} and \citet{schaefer2020} indeed find negative correlations between SFE and N/O at a given metallicity, though it should be noted that \citet{schaefer2020} use dust extinction to estimate local gas mass rather than measuring the gas mass directly. 

\rev{In the case of our high-D4000 spaxels, we do not find galactic winds to be a satisfactory explanation for their high N/O ratios. These spaxels are typically found at relatively low radii, and there appears to be little connection between N/O and radius at the high-metallicity end of our sample (see \autoref{no-oh-maps}, bottom right panel). These spaxels are also relatively old and would hence be expected to be comparatively metal-rich, given the known correlations between stellar ages and gas metallicities in galaxies \citep[e.g.][]{sm2020}; as such, we would expect infall of metal-rich gas to have relatively little impact compared to similar infalls on galaxies' metal-poor outer parts. A reduction in SFE, on the other hand, is a possible factor behind high-D4000 spaxels' behaviour: SFE and N/O remain negatively correlated at high metallicities \citet{schaefer2022}, and resolved CO maps of green valley MaNGA galaxies have shown then to possess suppressed SFE compared to ordinary star-forming galaxies \citep{brownson2020}.}

\rev{An additional possibility is that the highest-D4000 regions in our sample are experiencing infalls of metal-poor gas, with their present SFRs (as traced by the H$\alpha$ emission) insufficient  to re-enrich the ISM afterwards. It is possible that these high-D4000 regions had previously quenched and then been rejuvenated recently, which carries an implied sample-selection effect: rejuvenation of an old galaxy region via metal-poor inflows would allow such a region to be selected for inclusion in our sample, while an old region without such infall would not experience star-formation and so would not be included in our sample. \citet{belfiore2015} have previously proposed metal-poor gas infalls to explain individual galaxies displaying regions of elevated N/O and old stellar populations. This explanation is also similar to what \citet{yates2012} suggest for massive galaxies with low gas metallicities, based on results from semi-analytic modelling.}

At lower metallicities, significant N/O elevations appear to be associated with large galactocentric radii, which has previously been reported for ALM spaxels by \citet{luo2021} and which is entirely consistent with other aforementioned MaNGA work \citep{belfiore2017,schaefer2022}. However, the mass dependence (\autoref{no-oh-maps}, top right panel) remains important to consider here: at a given metallicity, more nitrogen-rich spaxels will be found in more massive galaxies on average and \textit{also} at larger galactocentric radii on average. SFE variations and galactic winds both provide plausible explanations for N/O variations at lower metallicities, with metal-poor inflows having the potential to further offset the N/O abundance by reducing O/H without a corresponding reduction to N/O.

\section{Summary \& Conclusion}\label{conclusion}

In this work, we have employed MaNGA data to assess the connection between local gas-phase abundances and local star-formation histories. We employed the PCA-based SFH classfications described for MaNGA in \citet{rowlands2018}, along with measurements of local emission line fluxes and local stellar properties. We considered spaxels with SFH classifications of star-forming, starburst, post-starburst and green valley; we then considered their behaviour on the N/O--O/H plane, while also considering how other stellar properties vary in N/O--O/H space. All spaxels, we reiterate, contain evidence of ongoing star-formation and would be classified as star-forming from emission line diagnostics alone.

We found that starburst regions possess diluted abundances when compared to ordinary star-forming regions, consistent with previous work. We also found post-starburst regions to behave very similarly to star-forming regions in N/O--O/H space, supporting a view in which gas is enriched rapidly following a starburst event; such a view is consistent with the stellar populations observed within MaNGA post-starburst regions, in which young stars are typically found to be significantly more metal-rich than older stars \citep{leung2023}. Furthermore, we found green-valley regions to posses elevated N/O abundances on average for their metallicity, which appears to relate to the behaviour of high-D4000 galaxy regions in our sample more generally. We argued that such regions \rev{may} have experienced recent metal-poor inflow and that they were possibly rejuvenated as a consequence\rev{, with SFE variations being another possible factor}. We also found various stellar population parameters ($\Sigma_*$, $\mathrm{M_\star}$, D4000) along with galactocentric radius to trend strongly with both O/H and N/O, with neither gas abundance parameter being obviously more or less informative than the other for tracing the evolution of galaxy regions.

Though various works (including this one) support a tight connection between gas-phase abundance and SFH, our results imply that gas-phase abundances are \textit{insensitive} to the precise form of the SFH at late times: both bursty SFHs and more gradual SFHs ultimately yield abundances that follow the same N/O--O/H relation. This finding might help to explain why certain longer-term stellar population parameters (such as $\Sigma_*$ and D4000) are so predictive of local gas metallicities: older and denser galaxy regions naturally have more time and opportunity to enrich their gas, with the precise form of the SFH having little impact on the final outcome.

\section*{Acknowledgements}

\rev{The authors thank the anonymous referee for their thoughtful remarks, which we feel to have improved our manuscript.} For the purpose of open access, the author has applied a Creative Commons Attribution (CC BY) licence to any Author Accepted Manuscript version arising. The support and resources from the Center for High Performance Computing at the University of Utah are gratefully acknowledged. \rev{The authors thank Sebastian Sanchez for useful feedback on an earlier draft of the manuscript.} NFB and VW acknowledge Science and Technologies Facilities Council (STFC) grant ST/V000861/1. NVA and VW acknowledge the Royal Society and the Newton Fund via the award of a Royal Society--Newton Advanced Fellowship (grant NAF\textbackslash{}R1\textbackslash{}180403). NVA acknowledges support from Conselho Nacional de Desenvolvimento Cient\'{i}fico e Tecnol\'{o}gico (CNPq). YL acknowledges support from Space Telescope Science Institute Director's Discretionary Research Fund grant D0101.90281, and SOFIA grant \#08-0226 (PI: Petric).

 Funding for the Sloan Digital Sky Survey IV has been provided by the Alfred P. Sloan Foundation, the U.S. Department of Energy Office of Science, and the Participating Institutions. SDSS-IV acknowledges support and resources from the Center for High-Performance Computing at the University of Utah. The SDSS web site is \url{www.sdss.org}. 

SDSS-IV is managed by the Astrophysical Research Consortium for the Participating Institutions of the SDSS Collaboration including the Brazilian Participation Group, the Carnegie Institution for Science, Carnegie Mellon University, the Chilean Participation Group, the French Participation Group, Harvard-Smithsonian Center for Astrophysics, Instituto de Astrof\'isica de Canarias, The Johns Hopkins University, Kavli Institute for the Physics and Mathematics of the Universe (IPMU) / University of Tokyo, Lawrence Berkeley National Laboratory, Leibniz Institut f\"ur Astrophysik Potsdam (AIP),  Max-Planck-Institut f\"ur Astronomie (MPIA Heidelberg), Max-Planck-Institut f\"ur Astrophysik (MPA Garching), Max-Planck-Institut f\"ur Extraterrestrische Physik (MPE), National Astronomical Observatories of China, New Mexico State University, New York University, University of Notre Dame, Observat\'ario Nacional / MCTI, The Ohio State University, Pennsylvania State University, Shanghai Astronomical Observatory, United Kingdom Participation Group, Universidad Nacional Aut\'onoma de M\'exico, University of Arizona, University of Colorado Boulder, University of Oxford, University of Portsmouth, University of Utah, University of Virginia, University of Washington, University of Wisconsin, Vanderbilt University, and Yale University.

\section*{Data Availability}

All non-MaNGA data used here are publically available, as are all MaNGA data as of SDSS DR17. The PCA code used in this analysis is stored at \url{https://github.com/KateRowlands/MaNGA-PCA}, and maps produced by the code (including those used in this article) are located at \url{https://doi.org/10.5281/zenodo.8277393}.

\bibliographystyle{mnras}
\bibliography{bibliography}

\begin{appendix}

\section{PCA boundaries}\label{pcabounds}

\begin{figure}
\begin{center}
	\includegraphics[trim = 2cm 0cm 1cm 0cm,scale=0.6]{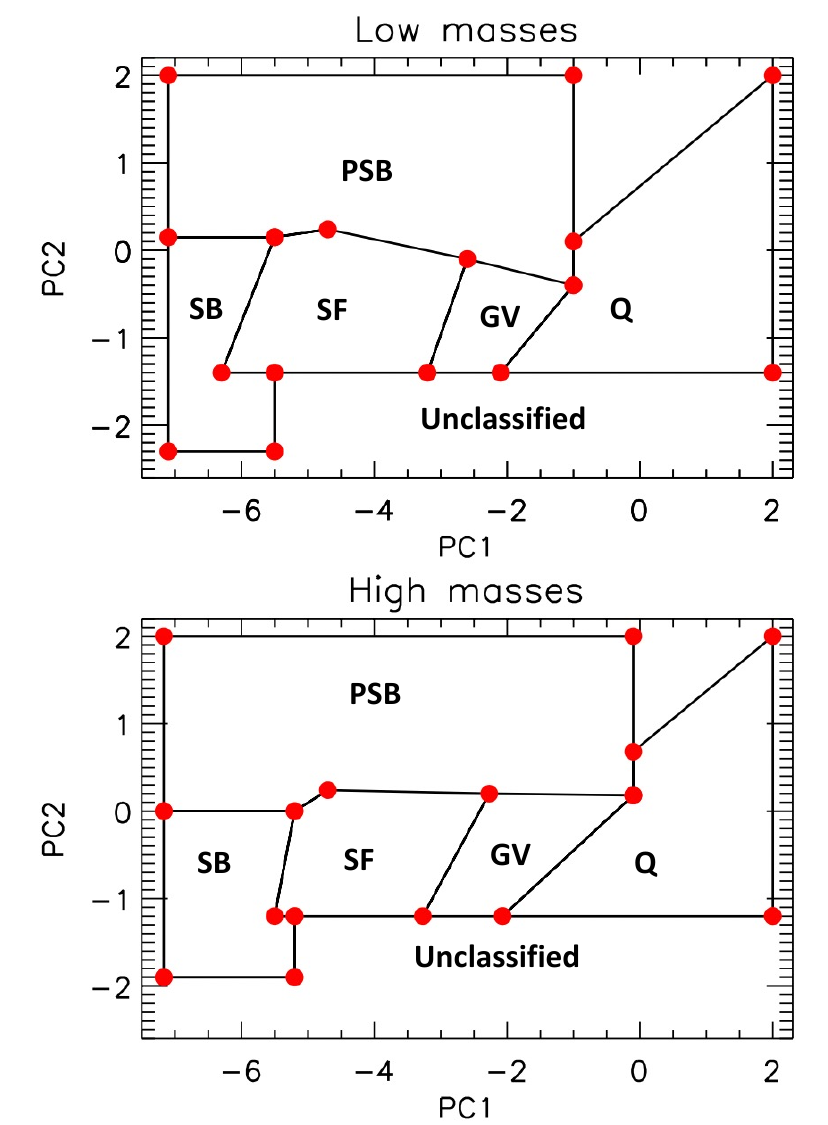} 
	\caption{\rev{Full PCA classification boundaries with defining vertices displayed as red circles.}}
	\label{boundaryfig_demo}
	\end{center}
\end{figure}

\rev{We parameterize the PCA boundaries with a series of vertices, as indicated in \autoref{boundaryfig_demo}; a small number of spaxels fall below these boundaries and are considered unclassified, as indicated in the main paper text. We employ different sets of vertices for NSA stellar masses below and above $10^{10}$ $\mathrm{M_\odot}$ with the default NSA cosmology ($h = 1, \Omega_M = 0.3, \Omega_\Lambda = 0.7$) employed in this case only. We present the vertex locations for lower-mass galaxies in \autoref{tablelow}, and we present the vertex locations for higher-mass galaxies in \autoref{tablehigh}.}

\begin{table*}
\begin{center}
\begin{tabular}{c|c c c c c c}
Classification &  &  & Vertices & & & \\
\hline
\hline
Star-forming (SF) & (-6.3, -1.4) & (-5.5, 0.15) & (-4.7, 0.24) & (-2.6, -0.1) & (-3.2, -1.4) \\
Starburst (SB) & (-7.1, -2.3) & (-7.1, 0.15) & (-5.5, 0.15) & (-6.3, -1.4) & (-5.5, -1.4) & (-5.5, -2.3) \\
Post-starburst (PSB) & (-4.7, 0.24) & (-5.5, 0.15) & (-7.1, 0.15) & (-7.1, 2) & (-1, 2) & (-1, -0.4)\\
Green valley (GV) & (-3.2, -1.4) & (-2.6, -0.1) & (-1, -0.4) & (-2.1, -1.4) \\
Quiscent (Q) & (-2.1, -1.4) & (-1, -0.4) & (-1, 0.1) & (2, 2) & (2, -1.4) \\
\end{tabular}
\end{center}
\caption{\rev{PCA boundaries' vertices in (PC1, PC2) space for galaxies with NSA stellar masses below $10^{10}$ $\mathrm{M_\odot}$ when using the default NSA cosmology.}}
\label{tablelow}
\end{table*}

\begin{table*}
\begin{center}
\begin{tabular}{c|c c c c c c}
Classification &  &  & Vertices & & & \\
\hline
\hline
Star-forming (SF) & (-5.5, -1.2) & (-5.2, 0.) & (-4.7, 0.24) & (-2.27, 0.2) & (-3.27, -1.2) \\
Starburst (SB) & (-7.17, -1.9) & (-7.17, 0) & (-5.2, 0) & (-5.5, -1.2) & (-5.2, -1.2) & (-5.2, -1.9) \\
Post-starburst (PSB) & (-4.7, 0.24) & (-5.2, 0) & (-7.17, 0) & (-7.17, 2) & (-0.1, 2) & (-0.1, 0.18) \\
Green valley (GV) & (-3.27, -1.2) & (-2.27, 0.2) & (-0.1, 0.18) & (-2.07, -1.2) \\
Quiscent (Q) & (-2.07, -1.2) & (-0.1, 0.18) & (-0.1, 0.68) & (2, 2) & (2, -1.2) \\
\end{tabular}
\end{center}
\caption{\rev{PCA boundaries' vertices in (PC1, PC2) space for galaxies with NSA stellar masses above $10^{10}$ $\mathrm{M_\odot}$ when using the default NSA cosmology.}}
\label{tablehigh}
\end{table*}

\end{appendix}
\label{lastpage}
\end{document}